\documentclass[twocolumn,aps,pre,floatfix,superscriptaddress]{revtex4-1}
\usepackage{graphics}
\usepackage{graphicx}
\usepackage{amsmath}
\usepackage{amssymb}
\usepackage{ulem}
\usepackage{color}

\begin{document}
\title{Analyzing Capture Zone Distributions (CZD) in Growth: Theory and Applications}
\author{T.L. Einstein}
\email[]{einstein@umd.edu}
\affiliation{Department of Physics, University of Maryland, College Park, Maryland 20742-4111, USA}
\affiliation{Condensed Matter Theory Center, University of Maryland, College Park, Maryland 20742-4111, USA}
\author{Alberto Pimpinelli}
\email[]{ap19@rice.edu}
\affiliation{Department of Physics, University of Maryland, College Park, Maryland 20742-4111, USA}
\affiliation{Rice Quantum Institute, Rice University, Houston, Texas 77005 USA}
\author{Diego Luis Gonz\'alez}
\email[]{diego.luis.gonzalez@correounivalle.edu.co}
\affiliation{Department of Physics, University of Maryland, College Park, Maryland 20742-4111, USA}
\affiliation{Departamento de F\'isica, Universidad del Valle, A.A. 25360, Cali, Colombia}
\date{\today}

\begin{abstract}
We have argued that the capture-zone distribution (CZD) in submonolayer growth can be well described by the generalized Wigner distribution (GWD) $P(s)=a s^\beta \exp(-b s^2)$, where $s$ is the CZ area divided by its average value.  This approach offers arguably the best method to find the critical nucleus size $i$, since $\beta \approx i + 2$.  Various analytical and numerical investigations, which we discuss, show that the simple GWD expression is inadequate in the tails of the distribution, it does account well for the central regime $0.5 < s < 2$, where the data is sufficiently large to be reliably accessible experimentally. We summarize and catalog the many experiments in which this method has been applied.
\end{abstract}

\maketitle 

\section{Prelude}
\begin{figure}[h]
\begin{center}
\includegraphics[trim = 1cm 7cm 11cm 4cm, clip,width=9 cm]{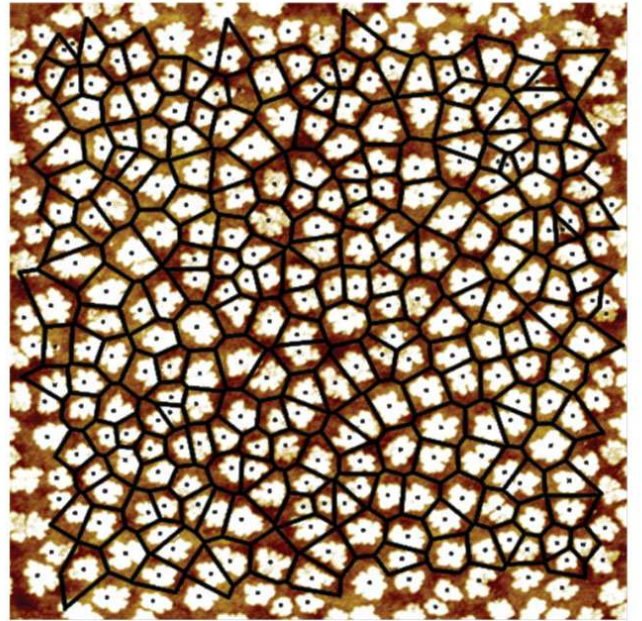}
\end{center}
\caption {An example (10 $\mu {\rm m} \times $ 10$\mu$m) AFM image of commercial pentacene. The island centers and Voronoi polygons are indicated by black dots and lines, respectively. From Ref.~\cite{CG08}.
}
\label{f:Voronoi}
\end{figure}

It is well established that a key goal to understanding the growth process is to find (in the early, aggregation regime) the size of the smallest stable cluster (denoted $i+1$, where $i$ is the size of the critical nucleus, the largest unstable cluster \cite{Sto70,V84,PV}).  While there are a few methods used traditionally to do so, more recent work has shown that it is particularly fruitful to consider the distribution of the area of capture zones \cite{BM95,BM96,ETB06,PE07}, i.e. Voronoi (proximity) cells constructed from the island centers. Cf.\ Fig.~\ref{f:Voronoi}. In particular, we shall see that this analysis provides information about the critical nucleus size $i$, i.e.\ the size of the largest island unstable to decay (so that the size of the smallest cluster that is assumed not to decay is $i+1$), a crucial ingredient in models of growth processes.  For random nucleation centers, i.e.\ Poisson-Voronoi (PV) diagrams, the capture-zone distribution (CZD) is expected to follow a gamma distribution ($\Gamma$D) \cite{K66,FN07},
\begin{equation}
\label{e:pGam}
P_{\Gamma}^{(\mathbf{\alpha)}}(s)=\frac{\alpha^{\alpha}}{\Gamma(\alpha)}s^{\alpha-1}e^{-\alpha \, s}.
\end{equation}

More generally we have argued \cite{PE07}, drawing from experiences analyzing the terrace-width distributions of vicinal surfaces \cite{EP99}, that the CZD is better described by the single-parameter generalized Wigner distribution (GWD):
\begin{equation}
\label{e:gwd}
P_\mathbf{\beta}(s) =a_\beta s^\beta \exp(-b_\beta s^2),
\end{equation}
\noindent where $s$ is the CZ area divided by its average value (so that $\langle s \rangle =1$); $a_\beta$ and $b_\beta$ are constants \cite{ab} that assure normalization and unit mean, respectively,  of $P(s)$. The derivation is based on a Fokker-Planck approach rooted in an overdamped Langevin analysis of cell size in which an external pressure from other cells hinders a cell from growing much larger than average while an entropic force impedes it from getting much smaller.  The entropic term is rooted in the mean-field assumption that the island nucleation rate is proportional to $n^{i+1}$, the product of the number $n$ of adatoms and the density of critical nuclei ($\propto n^i$) \cite{PExiv}, leading to the prediction that $\beta = i+1$ in two dimensions (2D) [and higher].  In 2D, which is the experimentally relevant case and so the focus of this short paper, this approximation turns out to be inadequate since nucleation occurs preferentially near CZ boundaries rather than uniformly \cite{LHE10,GE11}.  (It does work better in 1D, but that case requires---and permits---a more complicated analysis \cite{BM96,GPE11,Grin12}; v.i.)  As discussed in the following, more sophisticated analysis in 2D points to $\beta = i+2$ as better, consistent with most large-scale simulations and experiments.

\section{Refined Analysis and Simulations}
\begin{figure}[th]
\begin{center}
\includegraphics[width=8 cm]{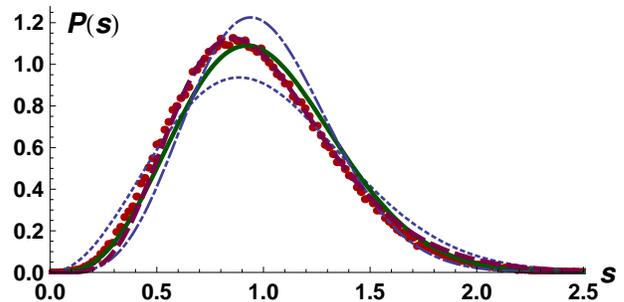}
\end{center}
\caption {Plots of Li, Han, and Evans's numerical data [red dots] \cite{LHE10}
for the CZD  for $i$ = 1 and $P_2(s)$
[dotted, blue line], $P_3(s)$ [solid, green line], and $P_4(s)$ [dash-dotted, blue
line], showing that $\beta = i+2 = 3$ does provide the best fit.  Also included is $P_{\Gamma}^{(7)}(s)$ [dashed, purple line]. From Ref.~\cite{PE10}.
}
\label{f:LHEfit}
\end{figure}

The need to go beyond our mean field Fokker-Planck analysis (and its key finding of $\beta = i+1$) \cite{PE07} were made evident by extensive simulations by Amar's \cite{SSA09} and Evans's \cite{LHE10} groups.  The former carried out kinetic Monte Carlo calculations of irreversible growth ($i = 1$) of point islands with dimensions $1-4$, for both square and triangular lattices in 2D and for two different point-island models.  In order to compute asymptotic behavior, they treat values of $R = 10^5 - 10^{10}$, where $R \equiv D/F$, $D$ the diffusion rate and $F$ the deposition flux.  They find better fits to CZDs with $\beta=3$ than with $\beta=2$.  The dependence on coverage (between 0.1 and 0.4 ML) is negligible, but there is some dependence on $R$ and on which of the two models of point islands is used. They also find better scaling (with $R$) of the peak height of the CZD using $\beta=3$, and comparable results for the two models.  Comparisons with some earlier calculations on extensive islands show similarities and differences, requiring further study to understand.   Li et al.~\cite{LHE10} considered both $i = 1$ and $i = 0$ at 0.1 ML.  As shown in Fig.~\ref{f:LHEfit}, $\beta=3$ accounts better for their $i = 1$ CZD curve than $\beta=2$.
While the GWD describes the CZD in the regime in which there is significant data in experiments (0.5  $< s <$ 2), it has shortcomings in the tails at both high and low $s$ \cite{LHE10,OA11}.  For large $s$, $P(s)$ may decay exponentially or like $s^{\beta{_\nu}} \exp(-A s^{-\nu})$ (with non-integer $\nu$ and $A$ some constant) rather than in Gaussian fashion.  Thus, in their short Comment, Li et al.~\cite{LHE10} plot the data in Fig.~\ref{f:LHEfit} on a log-log scale and get the best fit with $\nu = 1.5$ and $\beta_\nu \approx 4$ (and $\nu = 1.3$ and $\beta_\nu \approx 3$  for $i=0$); however, the relation between $\beta$ and $i$ should be determined by the central part of the CZD rather than power-law behavior for $s \ll 1$, so that $\beta_\nu$ and $\beta$ are expected to differ.  (Moreover, when fitting a particular curve, $\beta_\nu$ increases to ``compensate" decreasing $\nu$.) Their full report on this study is not yet published.

Most recently Oliveira and Aar\~ao Reis \cite{OA11} reported extensive simulations in 2D for point and extended (fractal and square) islands, with $i$ = 1 and 2, for $R = 10^6 - 10^{10}$.  Rather than using $s$ as their independent (scaling) variable, they choose $u = (x - \langle x \rangle)/\sigma_x \rightarrow (s-1)/[(\beta + 1)/(2b_\beta) -1]^{1/2}$, where $x$ is the number of lattice sites within a CZ, and $\sigma_x$ is the standard deviation; the final expression is for the GWD. This scaling procedure improves data collapse (for different values of $R$) by reducing the corrections to scaling for small $x$, where the continuum model underlying the derivation of Eq.~(\ref{e:gwd}) becomes inaccurate; on the other hand, this scaling hinders differentiating between values of $i$ (and so, by inference, $\beta$) in the central region in linear plots (but highlights differences in the tails in log-linear plots).  For point islands with $i=1$, both $\beta$ = 2 and 3 give "good fits" in the central (peak) region, with $\beta$ = 3 also adequate for small $s$ and neither doing well for large $s$; for $i=2$, $\beta$ = 3 gives a good fit of the center and large-$s$ tail.  For fractal and for square islands, there is good data for the various values of $R$, and $\beta = i +1$ gives a good fit in all 4 cases, while $\beta = i +2$ is not mentioned.  Typically there is Gaussian decay for large $s$.

While the numerical studies leave some open questions, the preponderance of evidence points to $\beta \approx i + 2$, which we can retrieve by refining our derivation \cite{PE10}: Noting that within a circular zone of radius $R$, the adatom density $n(r) \propto R^2 - r^2$, so that the integral over CZ area of $[n(r)]^{i+1} \propto R^{2i+4} \Rightarrow P(s) \propto s^{i+2}$. Figure~\ref{f:LHEfit} also shows that a gamma distribution might also be used to describe the numerical data, in this case with $\alpha = 7$.  (Indeed, Li et al.\ \cite{LHE10} noted that the data lies between this curve and $P_3(s)$.)  More generally, experimental data well described by the GWD with particular value of $\beta$ can also be fit with $P_{\Gamma}^{(\alpha)}(s)$, $\alpha \approx 2\beta + 1$ \cite{GC12}. (This approximation becomes progressively better for larger exponents [narrower distributions]: for $\beta > 7$, $2\beta + 1$ underestimates $\alpha$ by under 1\%; for $\beta > 3$ by $< 2$\%; but for $\beta$=1 by $\sim $ 7\%.)  However, the $\Gamma$D has no intrinsic connection with the critical nucleus size.

In a more detailed analysis based on a fragmentation model (FM) \cite{BM96}, we characterized systems in terms of two physically-rooted exponents, $\mathbf{\gamma}$ and $\mathbf{\delta}$ \cite{GE11}: (1) We took the probability to nucleate in a cell of size $s$ to be proportional to $s^\gamma P(s)$.  Ultimately, this implies that $P(s) \sim \exp(-A s^{-\gamma})$ (with $A$ some constant) for $s \gg 1$.  (2) We took the probability that a new center lies at a position \textbf{r} relative to the center of a preexisting cell is proportional to $|\mathbf{r}|^\delta$, assuming circular isotropy for simplicity.  The simplest case is the PV problem, for which $\gamma = 1$ and $\delta = 0$, with the second probability \textbf{r}-independent but $\propto s^{-1}$ \cite{GE11}.

As an analytically tractable example, we considered a point-island model with irreversible attachment in which zones are approximated as circular \cite{GE11}.  The isotropic solution of the appropriate steady-state diffusion equation gives an adatom density that increases from 0 at the interior island edge to a smooth maximum at the outer edge.  This leads to the deduction that $\gamma = 3$, which, with $\delta = 0$, accounts adequately for numerical data for $P(s)$, but less well for island density and the radial nearest-neighbor island probability distribution.  This deficiency can be remedied by taking $\delta = 1$  near the center and $\delta = 0$ near the edge \cite{GE11}.

\begin{figure}[t]
\begin{center}
\includegraphics[width=9 cm]{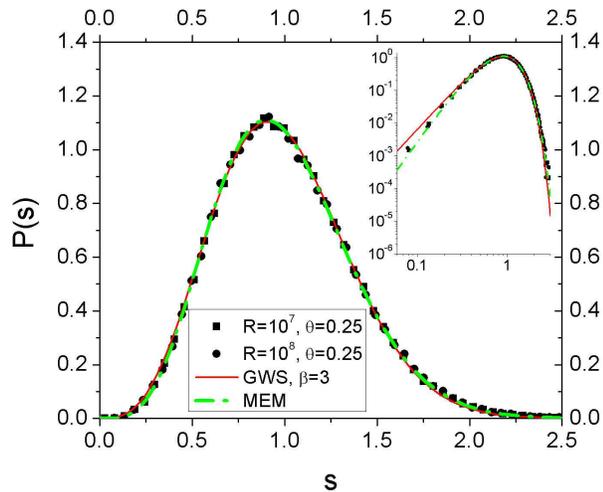}
\end{center}
\caption{Capture zone distribution in 2D with $i=1$. The GWS describes correctly the behavior of $P(s)$ for intermediate values of $s$. The maximum entropy method gives an excellent approximation for $P(s)$ even for large and small values of $s$, as seen in the replotting in the inset on a log-log scale.  See text.  From Ref.~\cite{GPE11}}.
\label{f:MEM}
\end{figure}

For $s \ll 1$, $P(s)$ is expected to behave like a power law in $s$, but the precise relation between this exponent and $i$ has not yet been determined, though, for $i=1$ good agreement with simulations have been found for $\beta = 4$ \cite{LHE10,GE11,OA11}.  In this regime $P(s)$ depends on the concentration of centers and ultimately, in the FM model, on $\delta$ \cite{GE11}. The skewness of the GWD also agrees well with numerical data in simulations \cite{OA11}.  However, trying to fit experimental data by just extracting the first few moments of the distribution were unsatisfactory \cite{GE00}.

Alternatively, we considered a maximum-entropy approach, noting that the first moment is unity by construction and the second moment, proportional to the product of island and monomer density, is also constant in the aggregation regime since latter cancels the former's $\theta^{1/3}$ behavior \cite{lam}, where $\theta$ is the coverage.  Thus, we obtained
\begin{equation}
\label{e:MEM}
P_{\rm MEM}(s)\approx A\,s^{\beta}\,e^{-B\,s^2-C\,s},
\end{equation}
\noindent where $A$, $B$, and $C$ are constants \cite{GPE11}.  As shown in Fig.~\ref{f:MEM}, the MEM expression accounts excellently for the numerical data for $i = 1$ for two different values of $R$, the ratio of the diffusion constant to the deposition rate.  In $P_{\rm MEM}(s)$ we set $\beta = 4$ consistent with the numerical results in Ref.~\cite{LHE10}.  In Table \ref{table01} we compare the quality of fits to these expressions. (See also Ref.~\cite{OA11}.)

\begin{table}[hbtp]
\begin{center}
    \begin{tabular}{ |@{\hspace{0.35em}}c @{\hspace{0.35em}}|@{\hspace{0.35em}}c @{\hspace{0.35em}} |@{\hspace{0.35em}} c @{\hspace{0.35em}}|@{\hspace{0.35em}} c @{\hspace{0.35em}}|@{\hspace{0.35em}}c @{\hspace{0.35em}} |@{\hspace{0.35em}} c @{\hspace{0.35em}}| @{\hspace{0.35em}}c @{\hspace{0.35em}}|}
    \hline
                  &  $\Gamma$D     & GWD-0    & GWD-1         & G$\Gamma$E       &   G$\Gamma$D   & MEM \\ \hline
     $10^3\,\chi^2$(all) & 3.010  & 1.660       & 1.726        & 0.402       & 0.334        & 0.518 \\ \hline
     $10^3\,\chi^2$(sig) & 1.722  & 0.826       & 0.873        & 0.381         & 0.287       & 0.294 \\ \hline
     $\nu$          &   1  & 2        & 2             & 1.5                      & 1.585        & NA \\ \hline
     $\beta$       &  6.277    & 3        & 3.065       & 4                    & 3.860          & 4 \\ \hline
    \end{tabular}
\end{center}
\caption{Values for the $\chi^2(all)$ and $\chi^2(sig)$ for four different analytical models of data for $i = 1$.  The values of $\chi^2$ are computed for the entire range of $s$ (all) and for just the range over which the data is significantly large (sig), viz.\ $0.5 < s < 2$.  For the first column, the gamma distribution ($\Gamma$D) $\beta$ is $alpha$ of Eq.~(\ref{e:pGam}).  GWD-0 has no free parameters, with $\beta = i+2 =3$, while GWD-1 lets $\beta$ vary to improve the fit.  In the column G$\Gamma$E the values of $\nu$ and $\beta$ are fixed at the values in Ref.~\cite{LHE10}, while in G$\Gamma$D (generalized gamma distribution) they are allowed to vary. MEM uses Eq.~(\ref{e:MEM}), with $A$, $B$, and $C$ fitted.  Adapted from Ref.~\cite{GPE11}}
\label{table01}
\end{table}

Another approach would be to allow two different values of $\gamma$: 2 for small $s$ and 1 for large $s$, mindful of the earlier result that $\gamma \approx (4+s)/(2+s)$ \cite{BE02}.  While such a "two-regime model" is not essential here, it is in the 1D case (where the $\gamma$ values are 4 and 3) \cite{GE11}.  More generally, there are several complications (e.g.\ the need for an integro-differential equation in the fragmentation analysis \cite{GPE11}) and more detailed analyses possible in 1D, leading to some controversies \cite{BM96,LHE10,GE11,GPE11,Grin12,SSA09,OA11} beyond the scope---and length limit---of this paper.

Two alternative approaches have long been used to gauge the critical nucleus size from experiment.  One is to measure the island size distribution (ISD) \cite{ETB06,lam,BE92} and then to fit it with the scaling formula (at least for $i$ = 1,2,3 and in 2D) \cite{AF95},
\begin{equation}
\label{e:Amar}
 f_i(u) = C_iu^i\exp(-i a_i u^{1/a_i}), \qquad \frac{\Gamma[(i+2)a_i]}{\Gamma[(i+1)a_i]} = (ia_i)^{a_i},
\end{equation}
\noindent with $u$ the island size divided by its mean and $C_i$ a normalization constant; Eq.~(\ref{e:Amar}) was deduced empirically from the expectations that (in the limit of large $R$) $f_i \sim u^i$ for small $u$, cut off exponentially for large $u$, and peak at $u=1$ (in marked contrast to the GWD, especially for small $\beta$).  While ISD cannot be expected to mimic the CZD in general \cite{ETB06,MR00}, Fanfoni et al.\ very recently presented some kinetic Monte Carlo (KMC) calculations on a simple model of quantum dot growth that point to a similarity between ISD and CZD at lower temperatures, when evolution of islands is dominated by atom motion along the periphery rather than attachment/detachment \cite{FA12}.  (However, they favor the $\Gamma$D in their fits.)  See Ref.~\cite{OA11} for recent results on the high-end tails.

Second, based on rate equation theory, it has long been known \cite{Sto70,V84} that at constant, relatively low temperature $T$, the density $N$ of [stable] islands (particularly the maximum density) satisfies the scaling relation \cite{V84,PJLP}
\begin{equation}
N \sim F^{\chi_i}, \qquad \chi_i^{\rm DLA} = i/(i+2) \; , \quad \chi_i^{\rm ALA} = 2i/(i+3)
\label{e:Ven}
\end{equation}
where the two relations of $\chi_i$ to $i$ are for diffusion-limited (DLA) and attachment-limited aggregation (ALA) regimes \cite{K97} in 2D, respectively.  There are many other regimes with signature relations for $\chi_i$ \cite{PJLP,PLJ}.  Also, the values in 3D differ, e.g.\ being $2i/(2i+5)$ for DLA (with compact islands and with no desorption).  In short, the value of $i$ deduced from $\chi_i$ depends strongly on the dominant mode of mass transport. In many cases one can characterize the temperature dependence by writing $N \sim (F/D)^{\chi_i}$, where $D$ has an activated, Arrhenius form, so that $N$ is expected to decrease rapidly with increasing $T$ \cite{PV}.

\section{Experimental Applications}

   \textbf{Pentacene on SiO$_2$} \cite{PB04}: Islands were fractal rather than compact/circular.  The CZD was found to depend on deposition rate.  We could fit the published data well with the GWD, with $\beta$ = 9 and 6 for high (1.5 nm/min) and low (0.15 nm/min) flux, respectively.  The ISD for high flux looked similar to the GWD, but for low flux the ISD was far broader and much more skewed.

    \textbf{Polar-conjugated molecule Alq$_3$ on passivated Si(100)} \cite{BGB02}: Brinkmann et al.\ fit their data with $\Gamma$D, quoting $\alpha = 10 \pm 2$.  Our fitting the areal data in their Fig.~6b is best with $\beta = 5$, lying between $\Gamma$D curves with $\alpha$ = 10 and 11. Mindful of Eq.~(\ref{e:Ven}) they measure $\chi_i = 2.00 \pm 0.05$.  While their interpretation that this implies $i$ = 5, inconsistent with our expectation (at least in 2D).  However, we can argue that the actual scaling regime differs from their assumption, leading to a compatible value of $i$.

    \textbf{Self-assembled Ge/Si(001) nanoislands} \cite{MMHI}: Displayed CZDs for 0.2, 0.8, and 1.0 ML on the wetting layer were are described by the GWD. Deduced values of $\beta$ every 0.2 ML rose slowly from about 2.4 at 0.2 ML to 4.68 at 1.0ML.  The associated value of $i$ was smaller than anticipated.  Island volume distributions fell from their initial maximum steadily to vanish before rising again to attain a smaller maximum.

   \textbf{InAs quantum dots on GaAs(001)} \cite{FA12,FP07}: In this example, nucleation is much faster than growth. The CZD of the quantum dots at 1.65 ML are well described by $P_{\Gamma}^{(4.1)}$, and $\alpha$ increases non-monotonically to 4.6 at 1.85 ML \cite{FP07}.  Fitting the published CZD at 1.65 ML with the GWD with $\beta = 2$ is comparably good, but Fanfoni \cite{Fpvt} reports that the fits with the $\Gamma$D are generally better and more appropriate because the nucleation sites are nearly random.  However, we achieved an even better fit using the FM with $\gamma = 1$ as for PV, but $\delta = 1$ rather than 0.  Fanfoni et al.'s analysis of the dot volume distribution is inconclusive \cite{FP07}.  Subsequently, Fanfoni et al.\ \cite{FA12} analyzed the CZD at 450$^\circ$C (1.74 ML, $F$ = 1.92 ML/min) and 528$^\circ$C (1.79 ML, $F$ = 1.68 ML/min), for which $\alpha$ values of 5.0 and 10.6, respectively, were extracted.  In the former case the volume distribution nearly coincided with the CZD but in the latter case it was much broader and more skewed.  Guided by the above-mentioned KMC simulations, they conclude that different mass-transport mechanisms dominate in the two regimes: periphery diffusion (consistent with PV) and attachment-detachment, respectively.

  \textbf{Metallic Ga droplets on GaAs(001)} \cite{NM12}: Fits of data at 2.9 ML to the GWD give $\beta$ = 6.9, 7.7, 8.6 for $T$ = 185, 190, 200$^\circ$C, corresponding to $i = 5 \pm 1, 6 \pm 1, 7 \pm 1$, consistent with values of $i$ for metals on semiconductors.  For smaller coverage 2.7 ML at 185$^\circ$C, $i = 4 \pm 1$, consistent (within error bars) with coverage-insensitive behavior.  For larger coverage (3.7 ML), ripening thwarts the analysis.

\textbf{Para-hexaphenyl (6P) films on amorphous mica} \cite{PTW11,TW12}: For 0.19 ML, $T$ = 300K, and $F$ = 0.04 ML/min on freshly cleaved mica, the needle-like islands have a CZD that is well (and best) fit with $\beta$ = 5
; the resulting $i$ = 3 is confirmed analysis of $N(F)$ data with Eq.~(\ref{e:Ven}) with $\chi_i = i/(i+2)$, yielding $i = 2.5 \pm 0.5$.  For such small $i$ they conclude that the 6P molecules must be lying on the surface, rather than standing up.  Different behavior is found at low $T$, attributable to different kinetics \cite{PTW11}.  Subsequent work on sputtered mica \cite{TW12} showed that the 6P molecules stand up, suggesting a higher $i$.  While no CZD analysis is given, values of $\chi_i$ in both the diffusion- and the attachment-limited regimes give $i = 7 \pm 2$.  ($i$ = 6 could correspond to a centered hexagon of standing rods [cf.\ Vienna sausage] as the smallest stable cluster.)

\textbf{6P on Ir(111)} \cite{H11}: The various analytic expressions do not describe the CZD data as well as other examples.  The best fit by the GWD is for $\beta = 1.6$, but the $s < 0.5$ data looks more like $\beta = 3$, and there are several points with unexpectedly high values for $s \lesssim 1$.

\textbf{Small admixtures of pentacenequinone (PnQ) with pentacene on ultrathin SiO$_2$ (UTO)} \cite{CG08}: While the above theoretical analyses assume deposition of single-species, study of CZD continues to provide information when there are impurities.  As the fraction of PnQ was increased at constant coverage (0.3 ML), the value of $\beta$ dropped from 6.7 below 1\% to 5.0 above 1\%, indicative of the poorer packing possible when PnQ was present.  For thick 50 ML films, this sudden change around 1\% is reflected in a sudden decrease grain size and a consequent decrease in linear mobility.

\textbf{C$_{60}$ on UTO} \cite{GC12}: The somewhat limited CZD data were fit comparably well with GWD and $\Gamma$D. The deduced values of $\beta$ had an inverted semicircular form between 298K and 483K, beginning and ending around 2.2 and peaking around 2.9 near 373K.  Over this range $N$ changed remarkably little, increasing slightly, then dipping a bit, in sharp contrast to the behavior expected in the comments after Eq.~(\ref{e:Ven}). Up to 3/4 of the atoms are monomers, but they are mostly immobile.  The most likely explanation of this is that surface defects act like impurities, confounding the simple scaling behavior.

\section{Further Applications}
In trying to understand unusual features in homoepitaxial growth on vicinal Cu(001), we concluded that a co-deposited impurity offered the best explanation \cite{HSPE11}.  Over a dozen possible elements were considered in a model involving two characteristic energies (lateral bond and diffusion barrier).  The elements fell into 4 classes, only one of which corresponded to the data.  In the course of these KMC simulations, we found different island morphologies in the 4 classes.  We generated CZDs for the various cases at coverages from 0.1 to 0.7 in steps of 0.1 \cite{HSPE11}.  The extracted values of $\beta$ likewise tended to divide into such classes, and for all but one class, $\beta$ increased with coverage, consistent with the experiments on Ge on Si(001) \cite{MMHI}.  The behavior of the exceptional class can be attributed to its repulsive nearest-neighbor interaction.

Finally, we have applied GZD analysis to various social phenomena. Examining the distribution of metro stations in central Paris, we find that the Voronoi distribution can be described by $P_{\Gamma}^{(\alpha = 8)}$, or better with the FM with $\gamma = 2$ and $\delta = 1.5$, indicative of an effective repulsion, greater than that between islands, consistent with the undesirability of having stations too close to each other \cite{GE11}.  Two decades ago it was observed that the secondary administrative divisions of France, the \textit{arrondissements} (districts), had properties of random cellular structures \cite{LCD93}.  Each has a chief town (corresponding to a county seat in the USA).  We constructed the distribution of the areas of these districts and also the areas of the Voronoi cells based on the chief towns.  These two distributions follow the same curve, which is well described by $P_{\Gamma}^{(\alpha \approx 4.4)}$, or the GWD with $\beta \approx 1.7$, or the FM with $\gamma = 1$ and $\delta = 0.5$ \cite{GE11,S09}.  Account for the pair correlation function, etc., of the chief towns requires inclusion of a hard-core circle, which turns out to optimally be 40\% of the mean radius \cite{GE11}. Furthermore, the GWD describes the area distributions of most other secondary administrative units, e.g.\ counties in the southeast USA ($\beta \approx 2$) and Polish powiaty ($\beta = 1.8 \pm 0.3$), as well as third-level rural gminy ($\beta = 2.0 \pm 0.4$) \cite{S09}.

Lastly we note that even the markings of giraffes have recently been considered from the perspective of random deposition but finite-thickness walls between Voronoi cells \cite{FZ12}.
\\*[4pt]

\section{Closing Summary Comments}
We have shown that the GWD provides an excellent accounting of CZDs in the region where the data in experiments is most reliable.   While the fits with the $\Gamma$D (or even a Gaussian, for large $\beta$) may also offer an adequate accounting, only the GWD offers a fundamental connection to the critical nucleus size: $\beta \approx i+2$. Further improvements in the theory, notably the fragmentation model, allow more detailed examination of tails and of other statistical functions. The approach applies to a much broader range of problems than just crystal growth.

\section*{Acknowledgments}
Work supported by UMD NSF-MRSEC Grant DMR 05-20471, NSF-CHE Grants 07-50334 and 13-05892.
We are grateful for collaborations on CZD analysis with theorists Rajesh Sathiyanarayanan and Ajmi BH. Hammouda, with the UMD experimental surface physics group, especially Ellen D. Williams, William G. Cullen, Brad Conrad, and Michelle Groce.

\end{document}